\documentclass[12pt]{article}

\usepackage{amsmath}

\usepackage{amssymb}
\usepackage{epsfig}

\usepackage{wrapfig}

\newcommand{\ope}{operator product expansion}
\newcommand{\pd}{\partial}

\newcommand{\np}[3]{\big[ #1 #2 \big]_{#3}}

\newcommand{\sub}[2]{{#1}_{\scriptscriptstyle #2}}

\begin{document}


\begin{titlepage}

\phantom{a}

\vfill

\begin{center}
{\LARGE \bf
Algebraic approach to parafermionic
conformal field theories \\
}
\vspace{20pt}
{
{ \large \bf Boris Noyvert
\\} 
\vspace{10pt}
{  e-mail: 
\hspace{-15pt}
\raisebox{-0.7pt}{
\includegraphics[height=8pt]{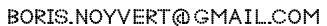}
}\\}
\vspace{3pt}
{ \small \it Department of Mathematics,}\\
{ \small \it  University of York,} \\
{ \small \it  YO10 5DD, York,  United Kingdom.}\\
}

\vspace{20pt}

\begin{abstract}

{\normalsize

Parafermionic conformal field theories
are considered on a purely algebraic basis.
The generalized Jacobi type identity is presented.
Systems of free fermions coupled to each other by nontrivial
parafermionic type relations are studied in detail.
A new  parafermionic conformal algebra
is introduced, it describes the $sl(2|1)_2/u(1)^2$
coset system.
}

\end{abstract}

\end{center}

\vfill

\end{titlepage}

\tableofcontents


\section{Introduction}


The first parafermionic two-dimensional conformal field theory
was introduced in 1985 in the classical article by V.~Fateev and
A.~Zamolodchikov \cite{Fateev:1985mm}.
Parafermion fields have fractional conformal dimension
and are not required to be local to each other,
the order of their mutual singularity can be any real number.
Parafermionic algebras can be seen as a generalization of
standard conformal chiral algebras
(vertex algebras in mathematical literature)
to the case of mutually nonlocal fields.

In that paper~\cite{Fateev:1985mm} the authors
study the $\mathbb{Z}_N$ invariant
parafermionic conformal field theory
with parafermion conformal dimensions being $\Delta_i=i(N-i)/N$
and
show that this
theory is equivalent
to the $sl(2)_N/u(1)$ coset.
Shortly after that in \cite{Zamolodchikov:1986gh}
the same authors used the $\mathbb{Z}_N$ parafermions
to build the minimal models of the $N=2$ superconformal algebra
and in \cite{Fateev:1985ig} they studied a new $\mathbb{Z}_3$
parafermionic theory generated by currents of dimension $4/3$.

In 1987 D.~Gepner \cite{Gepner:1987sm} introduced new parafermionic
algebras describing coset theo\-ries of type ${\mathfrak{g}}/u(1)^r$,
where ${\mathfrak{g}}$ is any simple Lie algebra and $r$ is its rank.
Conformal dimensions, fusion rules modular properties and partition
functions were found in this paper. However the exact algebraic
structure of the theory (e.g.~structure constants) remained unclear.

Later on parafermions (mainly the classical $\mathbb{Z}_N$ parafermions
from \cite{Fateev:1985mm}) were widely used in different areas
of conformal field theory and string theory.
A search for new $\mathbb{Z}_N$ invariant parafermionic
conformal field theories was a subject of a number of papers
during the last 15 years, we would like to mention
some of them.
Furlan et.al.~\cite{Furlan:1992va}
studied general $\mathbb{Z}_N$ invariant parafermionic theories
using the correlation function technique.
``Graded parafermions'' were introduced in \cite{Camino:1998pd},
these are based on the $osp(1|2)/u(1)$ coset construction.
Parafermions of the $A_2^{(2)}/u(1)$ coset were studied
in~\cite{Ding:2001ns}, their algebra is also graded
by the $\mathbb{Z}_N$ group.
Then in the recent years in series of papers
\cite{Dotsenko:2002gs, Dotsenko:2003zc, Dotsenko:2003kg, Dotsenko:2003ui, Dotsenko:2005se}
Dotsenko et.al.~studied
the second and the third solutions for the $\mathbb{Z}_N$
parafermion algebra from the Appendix to the paper~\cite{Fateev:1985mm}.
And finely last year Jacob and Mathieu \cite{Jacob:2005kc, Jacob:2005jz}
studied a new possibility for the $\mathbb{Z}_N$
algebra ($N$ even): the parafermion dimensions
being $\Delta_i=3i(N-i)/2N.$

A unified algebraic description of parafermionic
conformal field theories was still missing
in the physical literature.
The aim of this paper is to develop purely algebraic approach to
conformal algebras of parafermionic type,
and to illustrate its power on a few simple interesting examples.
The presented Jacobi-type condition
involves  only  \ope\ relations. This generalized Jacobi identity
in our opinion is the simplest tool to check the selfconsistence
of conformal algebras. It allows to use the so called
bootstrap approach in construction of new parafermionic theories,
the same approach which was used to build new W-algebras in the early
years of their study.

After the first version of this paper appeared on the web
we learned about the theory of generalized vertex algebras
developed in the mathematical literature. The most important reference
is
a monograph by
C.~Dong and J.~Lepowsky~\cite{Dong_Lepowsky}
\footnote{We are grateful to J.~Lepowsky for letting us know
about the monograph.}
in which conformal algebras of parafermionic type
(generalized vertex algebras)
are introduced for the first time from the mathematical point of view.
The first part of our paper (Sections~\ref{Framework and notation}
and~\ref{Generalized Jacobi identity}) has substantial overlap
with the theory developed in the book.
In particular the crucial notion of commutation factor and
the generalized Jacobi identity are present in~\cite{Dong_Lepowsky}.
There are a few levels of generalization in the monograph,
we suppose that our description of parafermionic conformal
algebras corresponds to the most general concept in the book
 - abelian intertwining algebra
(chapters 11, 12 of~\cite{Dong_Lepowsky}).

Mathematically oriented readers would also be interested to consult
a recent work by B.~Bakalov and V.~Kac~\cite{Bakalov_Kac}.
In this paper generalized vertex algebras are introduced using
the notion of polylocal fields. The definition
of a generalized vertex algebra here
is slightly more general than the one from~\cite{Dong_Lepowsky}.
The ``Borcherds identity'' (formula (27) in~\cite{Bakalov_Kac})
has exactly the same form as our
generalized Jacobi identity~(\ref{Jacobi})
up to a difference in the choice of normalization
of the commutation factors.

Contrary to the references~\cite{Dong_Lepowsky} and~\cite{Bakalov_Kac}
we focus here on the applications
of the algebraic approach to the study of parafermionic conformal field
theories. We employ this approach for the analysis of a few relatively simple
but instructive examples of parafermionic algebras,
for which we present the full set of algebraic relations,
calculate the structure constants and the commutation factors,
discuss the representation theory.

The paper has the following structure.
In Section~\ref{Framework and notation}
we fix notations, define conformal algebras of
parafermionic type and introduce the notion
of commutation factor which is crucial for
further developments and discuss consistency relations
for commutation factors.
In Section~\ref{Generalized Jacobi identity}
the Jacobi-type condition is derived.
The general commutator formula is stated in
Section~\ref{Commutator formula}. Next four sections
are devoted to examples of parafermionic algebras.
The first warmup example is
the classical $\mathbb{Z}_N$ invariant parafermion algebra.
Next we study the algebra of $sl(3)$ fermions,
formed by three dimension $1/2$ fields (associated with the
roots of the $sl(3)$ Lie algebra) coupled to each other
in nontrivial way. The third example is the generalization
of the second one to the $sl(n)$ case. And the last example
is the algebra of the $sl(2|1)_2/u(1)^2$ coset, this algebra
is generated by four dimension 1 fields and one free fermion field.
A short summary is given in Section~\ref{Summary}.


\section{Framework and notation}

\label{Framework and notation}


In two-dimensional conformal field theory
the conformal algebra 
is generated
by a set of conformal fields $\phi_i (z)$ of given conformal dimensions
$\Delta(\phi_i)$.
The algebraic relations between the fields are \ope s.
We refer the reader to the standard texts on conformal field theory
and vertex algebras
\cite{Belavin:1984vu, Ginsparg:1988ui, DiFrancesco:1997nk, Kac book, Lepowsky_Li}.

We will deal here with parafermionic conformal field theories,
in which the \ope\ of any two fields has the following form:
\begin{equation}
                                      \label{ope}
A(z) B(w) =\frac{1}{(z-w)^{\alpha}}
\Big(
\np{A}{B}{\alpha}(w)+\np{A}{B}{\alpha-1}(w)(z-w)
+\np{A}{B}{\alpha-2}(w)(z-w)^2+\cdots
\Big),
\end{equation}
i.e.~it is a general \ope\ with one important
restriction that
except the overall singularity $(z-w)^{-\alpha}$
the integer powers of $(z-w)$ only
are present in the brackets on the right hand side of the
equation. The singularity $\alpha=\alpha(A,B)$
depends on the fields $A$ and $B$, the dependence
is always assumed even when
not written explicitly.
This singularity can be calculated
from conformal dimensions as
\begin{equation}
\alpha=\Delta(A)+\Delta(B)-\Delta({\np{A}{B}{\alpha}}).
\end{equation}
The expression $\np{A}{B}{n}$ is the so called
$n$-product of fields $A$ and $B$. It is the field,
arising at the $(z-w)^{-n}$ term of the \ope\
of the fields $A(z)$ and $B(w)$ around $w$
as it appears in (\ref{ope}).
We usually assume that the most singular term $\np{A}{B}{\alpha}$
is not zero.

The singularity $\alpha$ doesn't have to be integer.
It can be any real number but normally it is rational.
The standard conformal
algebras (e.g.~Virasoro, affine Kac-Moody, W-algebras)
 are also of parafermionic type according to our
definition, but all the singularities are integer.
So we study here a more general theory, and
we will focus on the case when the \ope\ singularities
are not integer.


We should comment that parafermionic conformal field theories
are not the most general conformal field theories.
There are theories in which the powers of $(z-w)$
in the given \ope\ differ by a non-integer number.
The exact algebraic meaning of such relations is yet
to be understood.

In order to see that the algebra is
of parafermionic type one should choose an appropriate
basis of fields, since an \ope\ which looks like
(\ref{ope}) in one basis can be of mixed form
in another basis.

Now we want to introduce the mode expansions of the fields:
\begin{equation}
                                      \label{mode expansion}
A(z)=\sum_{n \in -\Delta(A)+\epsilon(A)+\mathbb{Z}}A_n z^{-n-\Delta(A)}.
\end{equation}
So we keep the standard notation for the modes, which is
very convenient from the physical point of view:
in this convention positive modes are annihilation operators
and negative modes are creation operators.
The sector
of the representation is encoded in the {\it{twisting}} $\ \epsilon \in \mathbb{R}/\mathbb{Z}$.
If $\epsilon(A)=0$ then the powers of $z$ are integer in (\ref{mode expansion})
and we say that
the field $A(z)$ is in the untwisted sector.
Not any value of the twisting is allowed, it should be consistent
with the \ope\ relations.
We should note that in the case of parafermionic algebras the action
of a field mode on a state changes the sector of the state.

The formula (\ref{mode expansion}) is reverted as following:
\begin{equation}
                                      \label{mode expansion reverted}
A_n=\oint_0 \mathrm{d}z \ A(z) z^{n+\Delta(A)-1},
\end{equation}
where $\oint_0$ means integration around zero, $\frac{1}{2\pi \mathrm{i}}$
is always assumed although is not written explicitly.

One should be able to express the reversed \ope\ $B(w)A(z)$
in terms of the \ope\ $A(z)B(w)$ itself.
In the case of standard conformal algebras
($\alpha\in \mathbb{Z}$) the rule is
$B(w)A(z)=-A(z)B(w)$ if both $A$ and $B$ are fermionic
and $B(w)A(z)=A(z)B(w)$ if at least one of them is bosonic.
When $\alpha\notin \mathbb{Z}$ it is not clear a priori
how to exchange the fields in the \ope , since some phases
are involved. We overcome this difficulty by multiplying
the \ope\ by its most singular term. So we introduce
the following axiom:
\begin{equation}
                             \label{mutual locality}
A(z)B(w)(z-w)^\alpha=
\mu_{\scriptscriptstyle A B}
B(w)A(z)(w-z)^\alpha.
\end{equation}
This equation is also a definition\footnote{
The definition of parafermionic mutual locality given in (8)
of~\cite{Bakalov_Kac} is essentially the same, the only difference
is in the normalization of the commutation factor.}
of the {\it commutation factor}
$\mu_{\scriptscriptstyle A B}$ which is a complex number
different from zero, and normally its absolute value is 1.

We add the axiom (\ref{mutual locality})
to our definition of the parafermionic conformal algebra.
This axiom is a key point of the theory,
once one accepts it, all the following derivations
are obtained almost automatically.

Now we are ready to clarify the exact meaning of an \ope\
(\ref{ope}). There are two approaches: mathematical and physical.
In the mathematical approach a field is a formal power series of
type~(\ref{mode expansion}), where $z$ is just a formal variable,
then the \ope\ of two fields is
just a product of two power series, but when one rearranges
some modes in the product, power terms in $(z-w)$ appear
(see e.g.~\cite{Kac book} for details).
In the physical approach the fields are physical quantum fields
in a two-dimensional quantum field theory,
and the variable is a complex variable, the real and imaginary parts
of which are physical coordinates. Then in the \ope\ one always
assumes the radial ordering (compare \cite{Ginsparg:1988ui}, equation (2.9)):
\begin{equation}
(z-w)^\alpha R(A(z) B(w))=\left\{
\begin{array}{ll}
(z-w)^\alpha A(z) B(w), &|z|>|w|,\\
\sub{\mu}{A B}(w-z)^\alpha B(w) A(z) , &|z|<|w|.
\end{array}
\right.
\end{equation}
One can use either of the approaches but sometimes one
is more convenient than another.

In the case of standard ($\alpha \in \mathbb{Z}$)
conformal algebras
$\mu_{\scriptscriptstyle A B}=-(-1)^\alpha$
if both $A$ and $B$ are fermionic and
$\mu_{\scriptscriptstyle A B}=(-1)^\alpha$
if at least one of them is bosonic.
But the commutation factor is different from
$\pm 1$ in general, and we will see such examples in the next sections.

By exchanging the fields in (\ref{mutual locality}) second time
one shows that the commutation factor should satisfy the following
consistency conditions:
\begin{equation}
\mu_{\scriptscriptstyle A B}
\mu_{\scriptscriptstyle B A}=1
\end{equation}
and if we assume that the term
$\np{A}{A}{\alpha_{\scriptscriptstyle A A}} \ne 0$
then it follows that
\begin{equation}
\mu_{\scriptscriptstyle A A}=1.
\end{equation}
However in some situations one should keep the possibility that
$\mu_{\scriptscriptstyle A A}=-1$, but if this is the case
then $\np{A}{A}{\alpha_{\scriptscriptstyle A A}} = 0$,
and then the real singularity exponent is actually
$\alpha_{\scriptscriptstyle A A}-1$
and not $\alpha_{\scriptscriptstyle A A}$.

If the \ope\ of two basic fields $B(w)$ and $C(v)$
gives a third one $D(v)$:
\begin{equation}
                          \label{B(w)C(v)}
B(w)C(v)=\frac{D(v)}{(w-v)^{\alpha_{\scriptscriptstyle B C}}}+ \cdots,
\end{equation}
then exchanging another basic field $A(z)$
with $B(w)$ and then with $C(v)$ is essentially the same
as exchanging $A(z)$ with $D(v)$. Therefore
$\sub{\mu}{A D}$ is proportional to $\sub{\mu}{A B} \sub{\mu}{A C}$.
The exact statement is
\begin{equation}
                                    \label{mumu}
\sub{\mu}{A B}\, \sub{\mu}{A C} =  \sub{\mu}{A D} \, \sub{r}{A B C},
\qquad
\sub{r}{A B C}=
(-1)^{\sub{\alpha}{A B}+\sub{\alpha}{A C}-\sub{\alpha}{A D}}
=\pm 1.
\end{equation}
It is also implicitly stated here that
$\sub{\alpha}{A B}+\sub{\alpha}{A C}-\sub{\alpha}{A D}\in \mathbb{Z}$.

To prove the statement consider the following regular function
in the 3 variables $z,w,v$:
\begin{equation}
R(z,w,v)=A(z) B(w) C(v) (z-w)^{\sub{\alpha}{A B}}
(z-v)^{\sub{\alpha}{A C}}(w-v)^{\sub{\alpha}{B C}}.
\end{equation}
Exchange the field $A(z)$ with $B(w)$ and then with $C(v)$
and then use the expansion~(\ref{B(w)C(v)}) and take
a limit $w\to v$, the result is
\begin{equation}
\lim_{w\to v}{R(z,w,v)}=
\sub{\mu}{A B} \sub{\mu}{A C} D(v) A(z)
(v-z)^{\sub{\alpha}{A B}+\sub{\alpha}{A C}}.
\end{equation}
Since this function should also be regular in $v$ around $z$
we conclude that
$\sub{\alpha}{A B}+\sub{\alpha}{A C}-\sub{\alpha}{A D}\in \mathbb{Z}$.
Now let us first use the expansion~(\ref{B(w)C(v)}) and then
exchange the fields $A(z)$ and $D(v)$ and then again take
the limit $w\to v$ to get:
\begin{equation}
\lim_{w\to v}{R(z,w,v)}=
\sub{\mu}{A D}  D(v) A(z)
(z-v)^{\sub{\alpha}{A B}+\sub{\alpha}{A C}-\sub{\alpha}{A D}}
(v-z)^{\sub{\alpha}{A D}}.
\end{equation}
Comparing the two expressions above we obtain
the statement~(\ref{mumu}).

The condition~(\ref{mumu}) also follows from the generalized
Jacobi identity stated in the next section
and should not be checked once all the Jacobi
identities are satisfied.

We should remark here that if
we do not specify any field in the \ope\
of $B(w)$ and $C(v)$, i.e.
\begin{equation}
B(w)C(v)=O((w-v)^{-\sub{\alpha}{B C}}),
\end{equation}
then there is no restriction on the product
$\sub{\mu}{A B} \sub{\mu}{A C}$ for any field $A$.

Conformal algebras of parafermionic type
are connected in some sense to the so called
color (or generalized, or ``$\epsilon$-'') Lie algebras
(\cite{Ree, Rittenberg:1978mr, Rittenberg:1978df, Scheunert:1978wn}
and some later papers).
These are Lie algebras with the bracket relation
$[x,y]=x y-\epsilon_{x y} y x$, so
the bracket is not symmetric or antisymmetric in general:
$[x,y]=-\epsilon_{x y}[y,x]$.


\section{Generalized Jacobi identity}

\label{Generalized Jacobi identity}


Here we derive the analogue of the Jacobi identities
for the conformal algebras of parafermionic type.
But first 
we would like to express
the $\np{B}{A}{n}$ products in terms of $\np{A}{B}{n}$
products.
According to our axiom (\ref{mutual locality})
the order of $A(z)$ and $B(w)$ in the \ope\ is reversed as following
\begin{equation}
                                      \label{ope new reversed}
B(w) A(z) =\frac{\mu_{\scriptscriptstyle B A}}{(w-z)^\alpha}
\Big(
\big[AB\big]_{\alpha}(w)+\big[AB\big]_{\alpha-1}(w)(z-w)+\big[AB\big]_{\alpha-2}(w)(z-w)^2+\cdots
\Big),
\end{equation}
and then one expands the fields in the second variable
(we also switch $w \leftrightarrow z$ to get the convenient form):
\begin{equation}
                                      \label{ope new reversed expanded}
\begin{aligned}
B(z) A(w) &=\frac{\mu_{\scriptscriptstyle B A}}{(z-w)^\alpha}
\Bigg(
\big[AB\big]_{\alpha}(w)+\bigg(\pd \big[AB\big]_{\alpha}(w)-\big[AB\big]_{\alpha-1}(w)\bigg)(z-w)
\\
&
+\bigg(\frac{1}{2!}\pd^2 \big[AB\big]_{\alpha}(w)
-\pd \big[AB\big]_{\alpha-1}(w)
+\big[AB\big]_{\alpha-2}(w)\bigg)(z-w)^2+\cdots
\Bigg),
\end{aligned}
\end{equation}
so
the $\big[BA\big]_{x}$ products are obtained from the $\big[AB\big]_{x}$ products as
\begin{equation}
                                      \label{BA products}
\big[BA\big]_{\alpha-n}=\mu_{\scriptscriptstyle B A}
\sum_{j=0}^n \frac{(-1)^j}{(n-j)!} \pd^{n-j}
\big[AB\big]_{\alpha-j}\, .
\end{equation}
It immediately follows that (if
$\big[A A\big]_{\alpha} \ne 0$)
\begin{equation}
\big[A A\big]_{\alpha-1}=\frac{1}{2} \pd \big[A A\big]_{\alpha},
\end{equation}
and more generally
\begin{equation}
\big[A A\big]_{\alpha-n}=\frac{1}{2} 
\sum_{j=0}^{n-1} \frac{(-1)^j}{(n-j)!} \pd^{n-j}
\big[A A\big]_{\alpha-j}\, ,
\qquad
n \in 2\mathbb{Z}+1,
\end{equation}
i.e. the pole $\big[A A\big]_{\alpha-n}$ for
odd $n$ is just a linear combination of derivatives
of higher poles.

Now we proceed to the Jacobi type condition
which involves three fields
$A$, $B$ and $C$. It is an analogue of the Jacobi
identity for Lie algebras. The condition is:
\begin{equation}
                                      \label{Jacobi}
\begin{aligned}
&\sum_{j \ge 0}
(-1)^j {\gamma_{\scriptscriptstyle A B} \choose j}
\big[A\big[B C\big]_{\gamma_{\scriptscriptstyle B C}+1+j}\big]
_{\gamma_{\scriptscriptstyle A B}+\gamma_{\scriptscriptstyle A C}+1-j}
\\
-
\mu_{\scriptscriptstyle A B}
(-1)^{\alpha_{\scriptscriptstyle A B}-\gamma_{\scriptscriptstyle A B}}
&\sum_{j\ge 0}
(-1)^j {\gamma_{\scriptscriptstyle A B} \choose j}
\big[B\big[A C\big]_{\gamma_{\scriptscriptstyle A C}+1+j}\big]
_{\gamma_{\scriptscriptstyle A B}+\gamma_{\scriptscriptstyle B C}+1-j}\\
=&
\sum_{j\ge 0}
 {\gamma_{\scriptscriptstyle A C} \choose j}
\big[\big[A B\big]_{\gamma_{\scriptscriptstyle A B}+1+j} C\big]
_{\gamma_{\scriptscriptstyle B C}+\gamma_{\scriptscriptstyle A C}+1-j}\, .
\end{aligned}
\end{equation}
All three sums are finite, the upper bound is given
by the order of singularity of the corresponding fields.
The parameters $\gamma$ differ from the corresponding singularity
exponents $\alpha$ by an integer number:
$\sub{\alpha}{A B}-\sub{\gamma}{A B},
\sub{\alpha}{A C}-\sub{\gamma}{A C},
\sub{\alpha}{B C}-\sub{\gamma}{B C} \in \mathbb{Z}$.
This identity is a modified version of the well known Borcherds
identity, however
an essential difference is the presence
of the commutation factor
$\mu_{\scriptscriptstyle A B}$.\footnote{
In mathematical literature a similar Jacobi type identity
is introduced in~\cite{Dong_Lepowsky}: 
the relation (11.9) in Chapter 11 ``Intertwining operators''
of the book contains the same information, although written
in a different form. The Borcherds identity (27) in
ref.~\cite{Bakalov_Kac} coincides 
with~(\ref{Jacobi}) above
if one uses the same normalization of the commutation factor.}

We say that the theory is associative
up to a certain
order in the expansion of two fields $A$ and $B$
if the condition (\ref{Jacobi}) is satisfied
for the inner $n$-products being of this and
more singular order. That means that if for any
fields $A$, $B$ one checks the identity
(\ref{Jacobi}) for
$\gamma_{\scriptscriptstyle A B} \ge n_{\scriptscriptstyle A B}-1$
then the theory is self consistent up to the term
$\np{A}{B}{n_{\scriptscriptstyle A B}}(w)(z-w)^{-n_{\scriptscriptstyle A B}}$
in the \ope\ of $A$ and $B$.

The generalized Jacobi identity~(\ref{Jacobi}) is derived using
the standard trick: double integrating
the function
\begin{equation}
f(z,w) = A(z) B(w) C(0)
(z-w)^{\gamma_{\scriptscriptstyle A B}}
w^{\gamma_{\scriptscriptstyle B C}}
z^{\gamma_{\scriptscriptstyle A C}}
\end{equation}
on $z$ and $w$ in two different ways: the first is
the integration  on $w$ around 0, then on $z$ around 0 minus
the opposite order; the second is the integration
on $z$ around $w$, then on $w$ around 0:
\begin{equation}
                            \label{double integration}
\oint_0 \mathrm{d}z \oint_0 \mathrm{d}w f(z,w) -
\oint_0 \mathrm{d}w \oint_0 \mathrm{d}z f(z,w)=
\oint_0 \mathrm{d}w \oint_w \mathrm{d}z f(z,w),
\end{equation}
there the internal (right) integral is taken first.
Then the first integration gives the left hand side of (\ref{Jacobi}),
the second integration gives the right hand side.

In the case of usual conformal algebras
(all the mutual singularities $\alpha$ are integer)
take $\gamma_{\scriptscriptstyle A B}=0$,
$\gamma_{\scriptscriptstyle B C}=p-1$, $\gamma_{\scriptscriptstyle A C}=q-1$,
where $p,q \in \mathbb{Z}$ to
get\footnote{One should use the identity ${0 \choose n}=\delta_{0,n}$ in the derivation.}
the Jacobi identity:
\begin{equation}
                                      \label{Jacobi simple}
\big[A\big[B C\big]_p\big]_q-
(-1)^{|A||B|}\big[B\big[A C\big]_q\big]_p=
\sum_{j=0}^\infty
 {q-1 \choose j}
\big[\big[A B\big]_{1+j} C\big]
_{p+q-1-j}\, ,
\end{equation}
which can be found elsewhere (see for example \cite{Thielemans:1994er}, formula (2.3.21)).
$|A|$ and $|B|$ are parities of the fields.
This formula is also valid when only $\alpha_{\scriptscriptstyle A B}$
is integer, and two other singularities are arbitrary.
($p$ and $q$ are not integer any more in this case.)



\section{Commutator formula}

\label{Commutator formula}


Here we derive the generalized commutator formula
for the modes of two fields, the \ope\ of which is
of parafermionic type.

In this section it will be convenient to change the notation.
We will consider here only one \ope\ relation:
\begin{equation}
                                      \label{ope_C}
A(z) B(w) =\frac{1}{(z-w)^\alpha}
\Big(
C^{(0)}(w)+C^{(1)}(w)(z-w)+C^{(2)}(w)(z-w)^2+\cdots
\Big).
\end{equation}
So comparing to (\ref{ope}) we have changed:
$C^{(j)}=\np{A}{B}{\alpha-j}\,$.

The main statement of this section is that
the \ope\ (\ref{ope_C}) is equivalent to the
following commutator formula for the field modes:
\begin{equation}
                                      \label{commutator formula}
\begin{aligned}
\sum_{j=0}^\infty
{\scriptstyle \alpha-k \choose  \scriptstyle j} (-1)^j A_m B_{n-m}&-& \hspace{-10pt}
\mu_{\scriptscriptstyle A B}(-1)^k
\sum_{j=0}^\infty
{\scriptstyle \alpha-k \choose  \scriptstyle j} (-1)^j B_{n-m'} A_{m'}&=
\sum_{l=0}^{k-1}
{\scriptstyle \gamma \choose \scriptstyle k-1-l} C^{(l)}_n .\\
\scriptstyle
m=-\Delta_A+\alpha-k+\gamma+1-j & &
\scriptstyle
\hfill m'=-\Delta_A+\gamma+1+j&
\end{aligned}
\end{equation}
Changing
$\gamma \to m+ \Delta_A-1, n \to n+m$
we get a different form which is usually more convenient:
\begin{equation}
                                      \label{commutator formula m+n}
\sum_{j=0}^\infty
{\scriptstyle \alpha-k \choose  \scriptstyle j}
(-1)^j
\Big(
A_{m+\alpha-k-j} B_{n-\alpha+k+j}- 
\mu_{\scriptscriptstyle A B}(-1)^k
B_{n-j} A_{m+j}
\Big)=
\sum_{l=0}^{k-1}
{\scriptstyle m+\Delta_A -1 \choose \scriptstyle k-1-l} C^{(l)}_{m+n}.
\end{equation}
Here
$k\in \mathbb{Z}$
is equal to the number of \ope\ singular terms
taken into account and can be any integer.
The derivation of the commutator formula
is based on the same standard trick:
double integration~(\ref{double integration}) of the following
expression
\begin{equation}
f(z,w)=A(z)B(w) (z-w)^{\alpha-k}(z/w)^\gamma w^{n+\Delta(C^{(0)})+k-2}
\end{equation}
in two ways as described in the previous section.
Then the first way of integration results in the left hand side
of (\ref{commutator formula}), the second way gives
the right hand side.

At this point it becomes clear why we call the factor $\mu$
a commutation
factor: it appears when one commutes two field modes.

Infinite sums are involved on the left hand side of commutator
formulas~(\ref{commutator formula},\ref{commutator formula m+n}).
However as usual the quadratic terms are ``nicely''
ordered: the modes with ``very large'' indices are on the right.
So if one acts by these infinite sum operators
on any state in the highest weight representation,
the sums are truncated and become finite.

In the case of usual conformal algebras
($\alpha \in \mathbb{Z}$) taking $k=\alpha$ one recovers
the standard commutator formula which can be found elsewhere.

Now we want to discuss how many terms in the \ope\
one should take and what are the relations between
the commutator formulas with different $k$.
The commutation relations (\ref{commutator formula m+n})
with smaller $k$ can always be obtained from those
with larger $k$. So the larger $k$ relations are more informative,
which is reasonable, since more terms in the \ope\ are taken
into account. The commutation relations are used to build
 representation theory of the algebra, e.g.~we want to be able
to exchange the modes in order to calculate the expectation
values of products of the modes. For that it is sufficient to know
all the singular terms (i.e.~of order $(z-w)^{-n},\ n>0$)
in the \ope\ of any two fields in the theory.
So for the needs of representation theory it is sufficient
to take $k$ a largest integer number smaller then the singularity
$\alpha$, smaller number of terms may be insufficient
to build a meaningful representation theory.






\section{Classical $\mathbb{Z}_N$ parafermions}

\label{Classical parafermions}


Here we present the classical example of parafermionic
conformal field theory, discovered in the pioneering work
\cite{Fateev:1985mm}. These theories are graded by
 $\mathbb{Z}_N$ group and coincide with the coset
\begin{equation}
\frac{sl(2)_N}{u(1)},
\end{equation}
the central charge of which is equal to
\begin{equation}
                           \label{c_Z_N}
c=2\,\frac{N-1}{N+2}.
\end{equation}

The algebraic data is the following:
there are $N$ fields $\psi_i$ of conformal dimensions
\begin{equation}
\Delta_i=\Delta_{N-i}=\frac{i(i-N)}{N},
\qquad
i=0,1,2,\ldots,N-1,
\end{equation}
the field $\psi_0$ is just the identity field.
The \ope s are $\mathbb{Z}_N$ graded:
\begin{equation}
                                \label{ope_Z_N}
\psi_i(z)\psi_j(w)=\frac{c_{i,j}}
{(z-w)^{\alpha_{i,j}}}
\Big(\psi_{i+j}(w)+O(z-w) \Big),
\end{equation}
where all the indices are taken modulo $N$
and the singularity is
\begin{equation}
\alpha_{i,j}=\Delta_i+\Delta_j-\Delta_{i+j}.
\end{equation}

The commutation factors are easily calculated
using the commutator factor relation~(\ref{mumu}),
they are all equal to 1:
\begin{equation}
\mu_{i,j}=1,
\end{equation}
and it follows that
\begin{equation}
c_{j,i}=c_{i,j}.
\end{equation}

One can check that the Jacobi identities (\ref{Jacobi})
are satisfied
up to the terms explicitly specified in the \ope s
(\ref{ope_Z_N})
if all the structure constants 
are fixed to the values, calculated in the paper \cite{Fateev:1985mm}
(using the method of correlation functions).
The results can be found there.
One also obtains from the Jacobi relations the next to leading term in
the expansion (\ref{ope_Z_N}):
\begin{equation}
\np{\psi_i}{\psi_j}{\alpha_{i,j}-1}=
\frac{\Delta_{i+j}+\Delta_i-\Delta_j}
{2\Delta_{i+j}}
\pd\psi_{i+j}.
\end{equation}
This term is also explicitly written in the Appendix~A of
\cite{Fateev:1985mm}.
If $i+j=N$ then this term vanishes. The next term in the \ope\
of two conjugate fields $\psi_i$ and $\psi_{N-i}$ has conformal
dimension 2, and we study from the Jacobi identities that
all these terms
$\np{\psi_i}{\psi_{N-i}}
{\alpha_{\scriptscriptstyle{i,{N-i}}}-2},\ 1\le i \le N/2$
are proportional to the same field. This dimension 2 field
can be identified with an energy--momentum field $T(z)$ of the theory.
So the ope\ of conjugate fields takes the form
\begin{equation}
                            \label{ope_Z_N_for_conjugate}
\psi_i(z) \psi_{N-i}(w) =\frac{1}{(z-w)^{2\Delta_i}}
\Big(
1+\frac{2\Delta_i}{c}T(w)(z-w)^2+O((z-w)^3) \Big),
\end{equation}
where we have fixed the normalization of the parafermionic fields:
$c_{i,N-i}=1,\ 0<i\le N/2$. The field $T$ is not an independent field
since it is equal to the nonsingular term in the \ope\
of $\psi_1(z)$ and $\psi_{N-1}(w)$. It satisfies the Virasoro algebra
of central charge $c$.

As a simple example of the use of the generalized Jacobi
identities~(\ref{Jacobi}) we show how
to calculate the value of the central charge.
We take $A=B=\psi_1, C=\psi_{N-1}$,
$\gamma_{\scriptscriptstyle A B}=\alpha_{1,1},
\gamma_{\scriptscriptstyle B C}=2\Delta_1-3,
\gamma_{\scriptscriptstyle A C}=2\Delta_1-1$
in (\ref{Jacobi}) and get the formula~(\ref{c_Z_N}).

Jacobi identities lead also to additional relations between
the next terms in the \ope s of basic fields. In particular
all the singular terms in the \ope s of basic fields
are related to the nonsingular terms in the \ope s
of other basic fields, which makes the representation
theory of the algebra reasonable.
There are also relations between nonsingular terms in the \ope s.

We would like to note here that other $\mathbb{Z}_N$ parafermionic
theories exist, some of them containing commutation factors
different from 1 (e.g.~the algebras in \cite{Camino:1998pd},
\cite{Ding:2001ns}, \cite{Jacob:2005jz}).


\section{$sl(3)$ fermions}

\label{$sl(3)$ fermions}


Here we discuss a very interesting example of a parafermionic conformal field
theory.
It is formed by 3 copies of standard free fermions,
coupled one to each other by
\begin{wrapfigure}[12]{l}[-18pt]{155pt}
\vspace{-12pt}
\includegraphics[width=150pt]{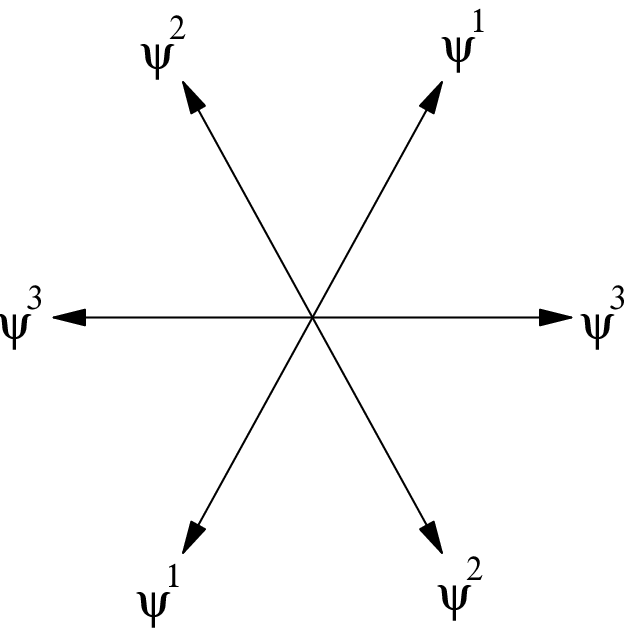}
\end{wrapfigure}
parafermionic type relations.
The theory is given by the following coset:
\begin{equation}
\frac{sl(3)_2}{u(1)^2},
\end{equation}
and so the central charge of this theory is
\begin{equation}
c=\frac{6}{5}.
\end{equation}
The theory is only a special case
of general coset construction
${\mathfrak{g}}/u(1)^r$
introduced and
studied in \cite{Gepner:1987sm}.
Here $\mathfrak{g}$ is a simple Lie algebra, $r$ is its rank.
However we are convinced that the theory deserves a special
study.

There are three fields of dimension $1/2$: $\psi^{(i)}, \ i=1,2,3$,
corresponding to the 3 pairs of opposite roots of $sl(3)$.
The \ope\ of each field with itself is the standard free fermion relation:
\begin{equation}
\psi^{(\alpha)}(z) \psi^{(\alpha)}(w)=\frac{1}{z-w}+O(z-w).
\end{equation}
The \ope\ of two different fields gives the third one:
\begin{equation}
\psi^{(\alpha)}(z) \psi^{(\beta)}(w)=
\frac{c_{\alpha, \beta} \psi^{(\gamma)}(w)}{(z-w)^{1/2}}+
O((z-w)^{1/2}),
\quad
\alpha \ne \beta \ne \gamma.
\end{equation}
The algebra is obviously $\mathbb{Z}_2 \times \mathbb{Z}_2$
graded. We will assign the following gradings to the fields:
$\psi^{(1)}$ grading is $(1,0)$, $\psi^{(2)}$ grading is $(0,1)$,
$\psi^{(3)}$ grading is $(1,1)$, and the identity field
is $(0,0)$ graded.

The structure constants are not all independent.
The axiom (\ref{mutual locality})
implies
\begin{equation}
c_{\alpha, \beta}=\mu_{\alpha, \beta} c_{\beta,\alpha} .
\end{equation}

There are 6 parameters
($\mu_{1,2}, \mu_{2,3}, \mu_{3,1},
c_{1,2}, c_{2,3}, c_{3,1}$)
to be fixed by the Jacobi identities (\ref{Jacobi}).
All the identities are satisfied if and only if
\begin{equation}
                               \label{sl3mu_c_solution}
\begin{aligned}
\mu_{1,2}=\mu_{2,3}= \mu_{3,1}=x^2,\\
c_{1,2}= c_{2,3}= c_{3,1}=\frac{x}{\sqrt{2}},
\end{aligned}
\qquad
x={\rm{e}}^{\pm\frac{{\rm{i}}\pi}{4}},
{\rm{e}}^{\pm\frac{3{\rm{i}}\pi}{4}}.
\end{equation}
We will make the choice $x={\rm{e}}^{-\frac{{\rm{i}}\pi}{4}}$ here,
then one obtains
the commutation factors:
$\mu_{1,2}=\mu_{2,3}= \mu_{3,1}=-{\rm i}=
-\mu_{2,1}=-\mu_{3,2}= -\mu_{1,3}$.

To calculate the relations between the modes of the fields
one has to use the commutator formula (\ref{commutator formula m+n}),
the result is:
\begin{align}
                                       \label{commutator_sl3_same}
\psi^{(\alpha)}_n \psi^{(\alpha)}_m +
\psi^{(\alpha)}_m \psi^{(\alpha)}_n&=
\delta_{0,m+n},\\
                                      \label{commutator_sl3_different}
\sum_{j=0}^\infty
{\scriptstyle j-1/2 \choose \scriptstyle j}
\Big(
{\rm{e}}^{\frac{{\rm{i}}\pi}{4}}
\psi^{(\alpha)}_{m-1/2-j} \psi^{(\beta)}_{n+1/2+j}
+{\rm{e}}^{-\frac{{\rm{i}}\pi}{4}}
\psi^{(\beta)}_{n-j} \psi^{(\alpha)}_{m+j}
\Big)
&=
\frac{1}{\sqrt{2}}\psi^{(\gamma)}_{m+n},
\end{align}
where $m,n \in \mathbb{Z}/2$ and
in the second relation $\alpha, \beta, \gamma$ are
all different and
ordered:
$(\alpha, \beta, \gamma)=(1,2,3)$ or $(2,3,1)$ or $(3,1,2)$.
The exact form of these commutation relations appears here
for the first time.

Unitarity can be introduced by the standard conjugation relation:
\begin{equation}
(\psi^{(\alpha)}_n)^\dag=\psi^{(\alpha)}_{-n},
\end{equation}
the generalized commutation relations above are invariant under this conjugation.

Next we would like to obtain the expression for the energy-momentum
field of the theory. We know the energy-momentum fields of the
three free fermion subalgebras:
$T^{(\alpha)}=-\frac{1}{2}\np{\psi^{(\alpha)}}{\pd \psi^{(\alpha)}}{0}
=\frac{1}{2}\np{\psi^{(\alpha)}}{\psi^{(\alpha)}}{-1}$.
Let us calculate the conformal weight of one of the parafermion fields
with respect to another energy-momentum field, i.e.~we want to calculate
$\np{T^{(\alpha)}}{\psi^{(\beta)}}{2}$ for $\alpha \ne \beta$.
Substitute in the generalized Jacobi identity~(\ref{Jacobi})
$A=B=\psi^{(\alpha)}, C=\psi^{(\beta)},
\gamma_{\scriptscriptstyle A B}=-2,
\gamma_{\scriptscriptstyle B C}=
\gamma_{\scriptscriptstyle A C}=-1/2$
to get that
\begin{equation}
\np{T^{(\alpha)}}{\psi^{(\beta)}}{2}=
\frac{1}{16} \psi^{(\beta)}, \quad \alpha \ne \beta.
\end{equation}
This result shows that two free fermions lie in the twisted
representation of the subalgebra generated by the third one.
It is easy to guess that the total energy-momentum tensor $T$
should be proportional to the sum of the three
energy-momentum fields. And the factor is easily calculated
from the condition
$\np{T}{\psi^{(\alpha)}}{2}=\frac{1}{2}\psi^{(\alpha)}$.
We conclude that the total energy-momentum field is
\begin{equation}
T=\frac{4}{5}
\sum_{\alpha=1}^3
T^{(\alpha)}=
-\frac{2}{5}
\sum_{\alpha=1}^3
{:}\psi^{(\alpha)}\pd\psi^{(\alpha)}{:},
\end{equation}
where ${:}\ {:}$ is the standard normal ordering.
We leave as an exercise to check that the field $T$ above
satisfies the \ope\ relation of the Virasoro algebra with
central charge $c=6/5$.

The three point correlation function of the parafermionic fields
is
\begin{equation}
{<}\psi^{(\alpha)}(z_1) \psi^{(\beta)}(z_2)
\psi^{(\gamma)}(z_3){>}
=
\frac{{\rm{e}}^{-\frac{{\rm{i}}\pi}{4}\epsilon_{\alpha \beta \gamma}}}
{(z_1-z_2)^{1/2} (z_2-z_3)^{1/2} (z_1-z_3)^{1/2}},
\end{equation}
where $\epsilon_{\alpha \beta \gamma}$ is the standard antisymmetric
tensor and we still follow the choice
$x={\rm{e}}^{-\frac{{\rm{i}}\pi}{4}}$.

We would like to say a few words about representation theory
of the $sl(3)$ fermion algebra. The main tool in the construction
of the representation theory is the commutation relations
(\ref{commutator_sl3_same}, \ref{commutator_sl3_different}).
A highest weight representation is constructed by applying
creation operators (nonpositive modes of the three generating
fermion fields) to a highest weight state. A highest weight state
is a state which is annihilated by all positive modes of the
generating fields. The highest weight state of the algebra is also
a highest weight state of the free fermion subalgebras.
The free fermion algebra has two highest weight representations:
the vacuum representation and the twisted representation.
So the highest weight representation of the full $sl(3)$ fermion algebra
should be a combination of these two representations.
It is easy to see from (\ref{commutator_sl3_same}, \ref{commutator_sl3_different})
that not all the combinations are allowed.
The consistent representations are build from the vacuum
highest weight state
${|}0,0,0{>}$, which is a vacuum state with respect to
all three subalgebras, and from the states
${|}0,\sigma,\sigma{>}$, ${|}\sigma,0,\sigma{>}$,
${|}\sigma,\sigma,0{>}$, which are vacuum states with respect to
one of the free fermion subalgebras and twisted states with respect to
two other subalgebras. The conformal dimension of the vacuum representation
is 0, the conformal dimension of the second type representation is
$4/5(1/16+1/16)=1/10$.

We would like to list here a few examples of states
in the vacuum representation:
\begin{equation}
\begin{aligned}
\psi^{(\beta)}_{-5/2}\psi^{(\alpha)}_{-3/2}\psi^{(\alpha)}_{-1/2}
&{|}0,0,0{>},\\
\psi^{(3)}_{3}\psi^{(1)}_{-4}\psi^{(3)}_{-5/2}
\psi^{(3)}_{-3/2}\psi^{(2)}_{1}\psi^{(1)}_{-3/2}
&{|}0,0,0{>},\\
\psi^{(2)}_{-1/2}\psi^{(1)}_{5/2}\psi^{(3)}_{-5}
\psi^{(1)}_{-3}\psi^{(1)}_{-2}\psi^{(2)}_{-7/2}
&{|}0,0,0{>}.
\end{aligned}
\end{equation}
The rule for the choice of integer or half-integer modes
is the following: as one adds operators from the left
at any stage the sum of modes should be integer,
if the $\mathbb{Z}_2 \times \mathbb{Z}_2$ charge of the string
of operators is $(0,0)$, and
the sum of modes should be half-integer otherwise.

The examples of states in the representation generated
from the highest weight state ${|}0,\sigma,\sigma{>}$
are
\begin{equation}
\begin{aligned}
\psi^{(3)}_{3/2}\psi^{(1)}_{-4}\psi^{(3)}_{-5}
\psi^{(3)}_{-3}\psi^{(2)}_{1/2}\psi^{(1)}_{-3/2}
&{|}0,\sigma,\sigma{>},\\
\psi^{(2)}_{-2}\psi^{(1)}_{5/2}\psi^{(3)}_{-5/2}
\psi^{(1)}_{-3}\psi^{(1)}_{-2}\psi^{(2)}_{-7}
&{|}0,\sigma,\sigma{>}.
\end{aligned}
\end{equation}
The rule is basically the same as above, the only difference
is that one should also count the state ${|}0,\sigma,\sigma{>}$
itself as it would carry a half-integer mode and a
$\mathbb{Z}_2 \times \mathbb{Z}_2$ grading $(1,0)$.

As a last remark on the representation theory we would like to note
that a kind of Poincare--Birkhoff--Witt (PBW) theorem should hold,
but we are even not sure
how to choose the PBW basis in the case of representations
discussed above.

A few words about parafermionic models of higher level cosets:
\begin{equation}
\frac{sl(3)_N}{u(1)^2}.
\end{equation}
The parafermionic algebra becomes $\mathbb{Z}_N \times \mathbb{Z}_N$
graded. It is generated by $N^2$ fields
$\psi^{(i,j)},\ i,j=0,1,\ldots,N-1$
with $\psi^{(0,0)}$ being the identity field.
The conformal dimensions are given by the following formula
\begin{equation}
\Delta_{i,j}=\max(i,j)-\frac{i^2+j^2-i\,j}{N}\,.
\end{equation}
The \ope\ of two fields $\psi^{(i,j)}$ and $\psi^{(k,l)}$
gives the field $\psi^{(i+k,j+l)}$, where all the indices
are taken modulo $N$. A study of generalized Jacobi identities
reveals that all the commutation factors are $2N$-roots of unity,
i.e.~$\mu^{2N}=1$,
and that some structure constants vanish if $N$ is odd.
These $\mathbb{Z}_N\times \mathbb{Z}_N$ graded algebras contain many
$\mathbb{Z}_N$ graded subalgebras, different from those
which have ever been studied.


\section{Simply laced fermions}

\label{$sl(n)$ fermions}


In this section we generalize the $sl(3)$
fermion algebra from the previous section.
The underlying root system is the root system
of a simple Lie algebra $\mathfrak{g}$ of A-D-E type.
The corresponding coset is
\begin{equation}
\frac{\mathfrak{g}_2}{u(1)^{r}},
\end{equation}
where $r$ is the rank of the algebra $\mathfrak{g}$.

The roots of a simply laced Lie algebra are all of the same
length. Two root directions may be either orthogonal or
having the angle $\pi/3$ between them.
We associate to every root direction $\alpha$ a free fermion algebra
generated by a field $\psi^{(\alpha)}$:
\begin{equation}
                              \label{free fermion simply laced}
\psi^{(\alpha)}(z) \psi^{(\alpha)}(w) =
\frac{1}{z-w}+O(z-w).
\end{equation}
Two fields are not coupled if the corresponding
root directions are orthogonal. If the root directions
are not orthogonal then the fields are coupled
in a parafermionic way:
\begin{equation}
\psi^{(\alpha)}(z) \psi^{(\beta)}(w)=
\left\{
\begin{array}{ll}
\scriptstyle
O((z-w)^0),
& \scriptstyle
\alpha \text{ and } \beta \text{ are orthogonal,}\\
\scriptstyle
\frac{c_{\alpha,\beta} \psi^{(\alpha+\beta)}(w)}
{(z-w)^{1/2}}+
O((z-w)^{1/2}),
& \scriptstyle
\alpha \text{ and } \beta \text{ are not orthogonal.}
\end{array}
\right.
\end{equation}
It is not clear what is the grading abelian group
in the case of algebraic relations described above.
Since due to~(\ref{free fermion simply laced})
the square of each element is identity,
the grading group should be $\mathbb{Z}_2^k$ for some integer $k$.
However the number of generating fields
in the theory is in general less than $2^k$.

The structure constants $c_{\alpha,\beta}$ and the
commutation factors $\mu_{\alpha,\beta}$ are to be fixed
by the Jacobi identities.
In fact there is a number of solutions to the Jacobi identities,
although we expect that all of them are equivalent from the physical
point of view.
If the root directions $\alpha$
and $\beta$ are not orthogonal, then
$\alpha, \beta, \alpha+\beta$ form an $sl(3)$ root subsystem,
and therefore the structure constants and the commutation factors
for the corresponding $sl(3)$ fermion subalgebra are subject to the
relations~(\ref{sl3mu_c_solution}). The commutation factors
between the orthogonal fermions are equal $\pm 1$.
The naive expectation that the fermions which are not coupled
are all mutually anticommutative is not true in general:
such solution exists only in the case $\mathfrak{g}=sl(4)$
and for the higher rank algebras the commutation factors
for the orthogonal root directions can not be chosen all of the
same sign!


We will present explicitly one of the solutions for the case
$\mathfrak{g}=sl(n)$, for general $n$.
The root directions of $sl(n)$ satisfy
the ``$so(n)$ pattern'' in the sense that the fusion
rules are identical to those of the $so(n)$ algebra
in the standard basis of antisymmetric $n \times n$ matrices.
So it would be convenient to label the fields by a pair
of two numbers:
$\psi^{(i,j)},\ 1<i \ne j <n$.
The order of indices is not important
($\psi^{(i,j)}\equiv\psi^{(j,i)}$),
so we will always assume that $i<j$.
In this two index notation the $sl(3)$ fermions
from the previous section
are written as
$\psi^{(1)}=\psi^{(2,3)}, \psi^{(2)}=\psi^{(1,3)},
\psi^{(3)}=\psi^{(1,2)}$.

If two fields have no common indices then they are not coupled.
If one of the indices is common for two root directions
then these root directions are not orthogonal and their sum
is labelled by the two distinct indices from the two pairs, for example
$(1,2)+(1,4)=(2,4)$ and the corresponding \ope\ is
\begin{equation}
\psi^{(1,2)}(z) \psi^{(1,4)}(w)
=\frac{c_{(1,2),(1,4)}
\, \psi^{(2,4)}}
{(z-w)^{1/2}}+
O((z-w)^{1/2}).
\end{equation}

We express the structure constants and the commutation factors
again in terms
of one variable $x$, which can take one of the four values
$x={\rm{e}}^{\pm\frac{{\rm{i}}\pi}{4}},
{\rm{e}}^{\pm\frac{3{\rm{i}}\pi}{4}}$.
In all the formulas below we assume that
$(i,j)<(k,l)$, that means
$i<k$ or $(i=k, j<l)$.
The structure constants are
\begin{equation}
c_{(i,j),(k,l)}=
\left\{
\begin{array}{ll}
x/\sqrt{2},& i=k \text{ or } j=l,\\
1/(x\sqrt{2}),& i=l \text{ or } j=k.
\end{array}
\right.
\end{equation}
If one index is common then the commutation factor is
\begin{equation}
\mu_{(i,j),(k,l)}=
\left\{
\begin{array}{ll}
x^2,& i=k \text{ or } j=l,\\
1/(x^2),& i=l \text{ or } j=k.
\end{array}
\right.
\end{equation}
If all the indices are different then
\begin{equation}
\mu_{(i,j),(k,l)}=
\epsilon_{i j k l}\ ,
\end{equation}
where $\epsilon_{i j k l}$
is antisymmetric in all 4 indices and
it is equal to 1 when $i < j < k <l$.
So we see, that the commutation factor
of two coupled fields is $\pm{\rm{i}}$,
and the commutation factor
of two fields which are not coupled is $\pm 1$.

To satisfy the Jacobi identities one should also
require the following condition
\begin{equation}
                               \label{NO_condition}
\np{\psi^{(i,j)}}{\psi^{(k,l)}}{0}+
{\rm{i}} \np{\psi^{(i,k)}}{\psi^{(j,l)}}{0}-
 \np{\psi^{(i,l)}}{\psi^{(j,k)}}{0}=0,
\quad i < j < k <l.
\end{equation}

The commutation relations read
\begin{align}
                                       \label{commutator_sln_same}
\psi^{(i,j)}_n \psi^{(i,j)}_m +
\psi^{(i,j)}_m \psi^{(i,j)}_n&=
\delta_{0,m+n},\\
\psi^{(i,j)}_n \psi^{(k,l)}_m
-\epsilon_{i j k l}
\psi^{(k,l)}_m \psi^{(i,j)}_n&=0,
\quad
i,j,k,l \text{ are all different,}
\end{align}
if the pairs of indices are the same or have no common indices.
If two pairs of indices have one common index then
the commutation relation between the corresponding field modes
is of parafermionic type:
\begin{equation}
                                      \label{commutator_sln_different}
\begin{aligned}
\sum_{s=0}^\infty
{\scriptstyle s-1/2 \choose \scriptstyle s}
\Big(
x^{-\epsilon}
\psi^{(i,j)}_{m-1/2-s} \psi^{(k,l)}_{n+1/2+s}
&+x^{\epsilon}
\psi^{(k,l)}_{n-s} \psi^{(i,j)}_{m+s}
\Big)
=
\frac{1}{\sqrt{2}}\psi^{(\bar i, \bar k)}_{m+n},\\
(i,j)<(k,l), \quad
\epsilon&=
\left\{
\begin{array}{ll}
1,& i=k \text{ or } j=l,\\
-1,& i=l \text{ or } j=k,
\end{array}
\right.
\end{aligned}
\end{equation}
where $\bar i$ and $\bar k$ are the two distinct indices
from the set $\{i,j,k,l\}$.
The condition (\ref{NO_condition}) is translated to
the following relation:
\begin{equation}
\sum_{n_1}
\psi^{(i,j)}_{m-n_1}\psi^{(k,l)}_{n_1}+
{\rm{i}}\sum_{n_2}
\psi^{(j,k)}_{m-n_2}\psi^{(j,l)}_{n_2}-
\sum_{n_3}
\psi^{(i,l)}_{m-n_3}\psi^{(j,k)}_{n_3}=0.
\end{equation}

%
%
%
%
%
%




\section{$sl(2|1)$ level 2 coset}

\label{$sl(2|1)$ level 2 coset}


In this section we introduce a new parafermionic conformal field
theory model.
\begin{wrapfigure}[7]{l}[-15pt]{150pt}
\vspace{-13pt}
\includegraphics[width=150pt]{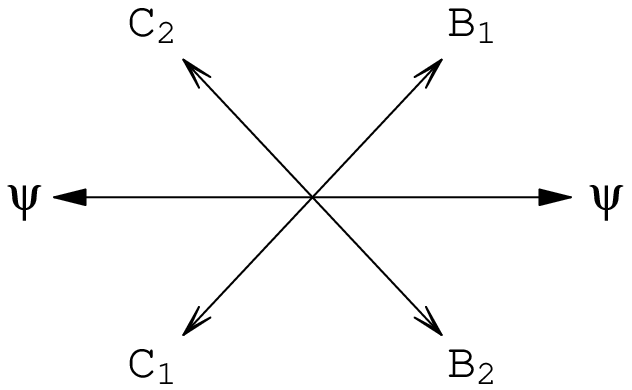}
\end{wrapfigure}
It is generated by one dimension $1/2$ free fermion field
and 4 dimension 1 fields.
In the end of the section we will show
that this model coincides with the coset of the $sl(2|1)$
affine superalgebra on level 2 by the two $u(1)$ currents, corresponding
to the Cartan elements of the $sl(2|1)$ algebra.

The fields are associated to the roots of the $sl(2|1)$ root system
as shown on the root diagram. The field $\psi$ is again the standard
free fermion field:
\begin{equation}
\psi(z) \psi(w)=\frac{1}{z-w}+O(z-w).
\end{equation}
Each pair of the dimension 1 fields, $(B_1,C_1)$ and $(B_2,C_2)$
forms the $psl(1|1)$ current algebra:
\begin{align}
B_i(z) C_i(w)&=\frac{1}{(z-w)^2}+O\big((z-w)^0\big),\\
B_i(z) B_i(w)&=O\big(z-w\big),\\
C_i(z) C_i(w)&=O\big(z-w\big),
\end{align}
where $i=1$ or $2$. The mutual commutation factor is
\begin{equation}
\sub{\mu}{B_i,C_i}=-1,
\end{equation}
which means that the fields are anticommutative and
therefore
\begin{equation}
C_i(z) B_i(w)=-\frac{1}{(z-w)^2}+O\big((z-w)^0\big).
\end{equation}

The root diagram tells us what should be the fusion rules
between the fields. The \ope\ of two fields associated
with the roots $\alpha$ and $\beta$ gives the field associated
with the root $\alpha+\beta$ if there is such a root, and
the \ope\ is not singular if $\alpha+\beta$
is not a root.
So the relations are
\begin{align}
                                       \label{B1B2}
B_1(z) B_2(w)&=\sub{\kappa}{B_1,B_2}
\frac{\psi(w)}{(z-w)^{3/2}}+O\big((z-w)^{-1/2}\big),\\
                                       \label{C1C2}
C_1(z) C_2(w)&= \sub{\kappa}{C_1,C_2}
\frac{\psi(w)}{(z-w)^{3/2}}+O\big((z-w)^{-1/2}\big),\\
B_1(z) C_2(w)&= O\big((z-w)^{1/2}\big),\\
C_1(z) B_2(w)&= O\big((z-w)^{1/2}\big),\\
\psi(z) B_1(w)&= \sub{\kappa}{\psi,B_1}
\frac{C_2(w)}{(z-w)^{1/2}}+O\big((z-w)^{1/2}\big),\\
\psi(z) C_2(w)&= \sub{\kappa}{\psi,C_2}
\frac{B_1(w)}{(z-w)^{1/2}}+O\big((z-w)^{1/2}\big),\\
\psi(z) B_2(w)&= \sub{\kappa}{\psi,B_2}
\frac{C_1(w)}{(z-w)^{1/2}}+O\big((z-w)^{1/2}\big),\\
\psi(z) C_1(w)&= \sub{\kappa}{\psi,C_1}
\frac{B_2(w)}{(z-w)^{1/2}}+O\big((z-w)^{1/2}\big),
\end{align}
where the coefficients $\kappa$
are structure constants.

We have checked the Jacobi identities to obtain
the commutation factors:
\begin{equation}
\sub{\mu}{B_1,B_2}=\sub{\mu}{C_1,C_2}
=\sub{\mu}{B_1,C_2}=\sub{\mu}{C_1,B_2}=
\sub{\mu}{\psi,B_2}=\sub{\mu}{\psi,C_2}=
-\sub{\mu}{\psi,B_1}=-\sub{\mu}{\psi,C_1}=-\mathrm{i},
\end{equation}
and the structure constants:
\begin{equation}
\sub{\kappa}{\psi,B_1}=\sub{\kappa}{\psi,C_1}=
-\sub{\kappa}{B_1,B_2}=\sub{\kappa}{C_1,C_2}=
\frac{\mathrm{e}^{-\mathrm{i}\pi/4}}{\sqrt{2}},
\quad
\sub{\kappa}{\psi,B_2}=\sub{\kappa}{\psi,C_2}=
\frac{\mathrm{e}^{\mathrm{i}\pi/4}}{\sqrt{2}}.
\end{equation}

In the relations above the following singular
terms are not specified:
$\np{B_1}{B_2}{1/2}$ and $\np{C_1}{C_2}{1/2}$.
One has to take these terms into account in order to be able
to write the meaningful generalized commutation relations
between $B_1$ and $B_2$ and between $C_1$ and $C_2$.
The Jacobi identities imply that these terms are proportional,
so only one new field has to be introduced. It has conformal dimension
$3/2$, we will call it $G$. The full form of
\ope s (\ref{B1B2}) and (\ref{C1C2}) is
\begin{align}
B_1(z) B_2(w)&=\sub{\kappa}{B_1,B_2}
\left(
\frac{\psi(w)}{(z-w)^{3/2}}
+
\frac{\frac{1}{2}\pd\psi(w)-
\mathrm{i} \frac{\sqrt{3}}{2} G(w)}{(z-w)^{1/2}}
\right)
+O\big((z-w)^{1/2}\big),\\
C_1(z) C_2(w)&= \sub{\kappa}{C_1,C_2}
\left(
\frac{\psi(w)}{(z-w)^{3/2}}
+
\frac{\frac{1}{2}\pd\psi(w)-
\mathrm{i} \frac{\sqrt{3}}{2} G(w)}{(z-w)^{1/2}}
\right)
+O\big((z-w)^{1/2}\big).
\end{align}
The \ope s of the field $G$ with other basic fields
read
\begin{align}
\psi(z) G(w)&=O\big((z-w)^{0}\big),  
\\
G(z)G(w)&=
\frac{- \frac{5}{3}}{(z-w)^3}+ \nonumber \\
&+\frac{-\frac{4}{3} \np{B_1}{C_1}{0}(w)
-\frac{4}{3} \np{B_2}{C_2}{0}(w)+\frac{1}{3}\np{\psi}{\psi}{-1}(w)
}{z-w}
+O\big((z-w)^{0}\big),\\
G(z)B_{1,2}(w)&=
\frac{
\frac{5}{2\sqrt{6}}\mathrm{e}^{\pm \mathrm{i}\pi/4} C_{2,1}(w)
}{(z-w)^{3/2}} + \nonumber \\
&+
\frac{
\sqrt{\frac{2}{3}} \mathrm{e}^{\pm \mathrm{i}\pi/4} \pd C_{2,1}(w)
\mp \frac{\mathrm{i}}{2\sqrt{3}} \np{\psi}{B_{1,2}}{-1/2}(w)
}{(z-w)^{1/2}}
+O\big((z-w)^{1/2}\big),\\
G(z)C_{1,2}(w)&=
\frac{
\frac{5}{2\sqrt{6}}\mathrm{e}^{\pm \mathrm{i}\pi/4} B_{2,1}(w)
}{(z-w)^{3/2}} + \nonumber \\
&+
\frac{
\sqrt{\frac{2}{3}} \mathrm{e}^{\pm \mathrm{i}\pi/4} \pd B_{2,1}(w)
\mp \frac{\mathrm{i}}{2\sqrt{3}} \np{\psi}{C_{1,2}}{-1/2}(w)
}{(z-w)^{1/2}}
+O\big((z-w)^{1/2}\big),
\end{align}
The commutation factors of $G$ are related to the commutation factors
of $\psi$:
\begin{equation}
\sub{\mu}{G,A}=-\sub{\mu}{\psi,A},
\end{equation}
where $A$ is any of the basic fields
$\psi, G, B_1, C_1, B_2, C_2$. In particular
$\sub{\mu}{\psi,G}=-1$, i.e.~the fields
$\psi$ and $G$ are anticommutative.

The Jacobi identities are satisfied modulo a null field
condition
\begin{equation}
\np{B_1}{C_1}{0}-\np{B_2}{C_2}{0}
-\mathrm{i}\frac{\sqrt{3}}{2}\np{\psi}{G}{0}=0.
\end{equation}

The algebra of 6 basic fields $\psi, B_1, C_1, B_2, C_2, G$
is closed in the sense that all the singular terms
in the \ope s of two basic fields are expressed in terms
of basic fields, their derivatives and composite fields.
By ``composite field'' we understand a field which is
equal to a nonsingular term in the \ope\
of two basic fields.

The algebra has a $\mathbb{Z}_2\times \mathbb{Z}_2$ grading
and a $U(1)$ charge. The fields $B_1, C_1$ have grading
$(0,1)$, $B_2, C_2$ - $(1,0)$, $\psi, G$ - $(1,1)$.
The fields $B_1$ and $C_2$ have charge $+1$,
$B_2$ and $C_1$ have charge $-1$, and $\psi, G$ - charge 0.

The energy-momentum field of the theory is
\begin{equation}
T=\frac{2}{3}
\left(\np{\psi}{\psi}{-1}-\np{B_1}{C_1}{0}-\np{B_2}{C_2}{0}
\right).
\end{equation}
It satisfies the Virasoro algebra of central charge
\begin{equation}
c=-2.
\end{equation}

The field $G(z)$ generates the $N=1$ superconformal algebra
of central charge $-5/2$, the corresponding Virasoro field
is $T_G=\np{G}{G}{1}/2$. The full Virasoro algebra decouples
to a sum of two commuting parts: $T=T_G+T_\psi$, there
$T_\psi=\np{\psi}{\psi}{-1}/2$
is the Virasoro field of the free fermion subalgebra.

The lengthy generalized commutation relations are
\begin{align}
&\psi_n \psi_m + \psi_m \psi_n=\delta_{n+m,0},\\
&(B_i)_n(C_i)_m+(C_i)_m (B_i)_n=n\delta_{n+m,0},
\\
&\sum_{j=0}^\infty
{\scriptstyle j-1/2 \choose \scriptstyle j}
\Big(
{\rm{e}}^{\pm\frac{{\rm{i}}\pi}{4}}
\psi_{m-1/2-j} (B_{1,2})_{n+1/2+j}
-{\rm{e}}^{\mp\frac{{\rm{i}}\pi}{4}}
(B_{1,2})_{n-j} \psi_{m+j}
\Big)
=
\frac{1}{\sqrt{2}}(C_{2,1})_{m+n},\\
&\sum_{j=0}^\infty
{\scriptstyle j-1/2 \choose \scriptstyle j}
\Big(
{\rm{e}}^{\pm\frac{{\rm{i}}\pi}{4}}
\psi_{m-1/2-j} (C_{1,2})_{n+1/2+j}
-{\rm{e}}^{\mp\frac{{\rm{i}}\pi}{4}}
(C_{1,2})_{n-j} \psi_{m+j}
\Big)
=
\frac{1}{\sqrt{2}}(B_{2,1})_{m+n},\\
&\sum_{j=0}^\infty
{\scriptstyle j-1/2 \choose \scriptstyle j}
\Big(
(B_{1})_{m-1/2-j} (C_{2})_{n+1/2+j}+
\mathrm{i} (C_{2})_{n-j}(B_{1})_{m+j}
\Big)
=0,\\
&\sum_{j=0}^\infty
{\scriptstyle j-1/2 \choose \scriptstyle j}
\Big(
(C_{1})_{m-1/2-j} (B_{2})_{n+1/2+j}+
\mathrm{i} (B_{2})_{n-j}(C_{1})_{m+j}
\Big)
=0,\\
&\sum_{j=0}^\infty
{\scriptstyle j-1/2 \choose \scriptstyle j}
\Big(
{\rm{e}}^{\frac{{\rm{i}}\pi}{4}}
(C_{1})_{m-1/2-j} (C_{2})_{n+1/2+j}-
{\rm{e}}^{-\frac{{\rm{i}}\pi}{4}}
(C_{2})_{n-j}(C_{1})_{m+j}
\Big)
=
\nonumber \\
&=
-\sum_{j=0}^\infty
{\scriptstyle j-1/2 \choose \scriptstyle j}
\Big(
{\rm{e}}^{\frac{{\rm{i}}\pi}{4}}
(B_{1})_{m-1/2-j} (B_{2})_{n+1/2+j}-
{\rm{e}}^{-\frac{{\rm{i}}\pi}{4}}
(B_{2})_{n-j}(B_{1})_{m+j}
\Big)
=
\nonumber \\
&=\frac{1}{2\sqrt{2}}
\left(
(m-n-\frac{1}{2})
\psi_{m+n}
-\mathrm{i} \sqrt{3} G_{m+n}
\right), \\
&\psi_n G_m + G_m \psi_n=0,
\end{align}
where we omit all the other generalized commutation relations
involving field $G$, since the expressions are too long.
The mode expansions of the composite fields should be calculated
using the same generalized commutator
formula~(\ref{commutator formula m+n}).

Now we want to show that the above algebra describes
the $sl(2|1)_2/u(1)^2$ coset. We introduce two commuting free bosons
$\phi_1$ and $\phi_2$:
\begin{equation}
\begin{aligned}
\phi_1(z) \phi_1(w)&=-\log(z-w),\\
\phi_2(z) \phi_2(w)&=\log(z-w).
\end{aligned}
\end{equation}
(Note the sign difference!)
Then the currents of the $sl(2|1)$ affine algebra
are expressed as
\begin{equation}
\begin{aligned}
H_1(z) &\sim \mathrm{i}\pd \phi_1(z),
&J^+(z) & \sim \psi(z) \mathrm{e}^{\mathrm{i}\phi_1(z)},\\
H_2(z) &\sim \mathrm{i}\pd \phi_2(z),
&J^-(z) & \sim \psi(z) \mathrm{e}^{-\mathrm{i}\phi_1(z)},\\
F_{1,2}^+(z) & \sim B_{1,2}(z)
\mathrm{e}^{\frac{\mathrm{i}}{2}(\phi_1(z)\pm\phi_2(z))},
&
{F}_{1,2}^-(z) & \sim C_{1,2}(z)
\mathrm{e}^{\frac{\mathrm{i}}{2}(-\phi_1(z)\pm\phi_2(z))}.
\end{aligned}
\end{equation}
There are 4 bosonic currents ($H_1, H_2, J^+, J^-$) and 4 fermionic
currents ($F_1^+, F_1^-, F_2^+, F_2^-$), they form an $sl(2|1)$
affine algebra on level 2. $H_1, H_2$ correspond to the two Cartan
elements of $sl(2|1)$, $J^+, J^-$ - to the two even roots,
$F_1^+, F_1^-, F_2^+, F_2^-$ - to the 4 odd roots.
The fields $H_1, J^+, J^-$ generate the $sl(2)$ affine subalgebra
on level 2.


\section{Summary}

\label{Summary}


We described conformal field theories of parafermionic type
from the algebraic point of view. The main algebraic tool,
the Jacobi type identity, is presented in the explicit form.
Using this identity we calculated the commutation factors and
the structure constants for the following cosets:
$sl(3)_2/u(1)^2$, $sl(N)_2/u(1)^2$, $sl(2|1)_2/u(1)^2$.
We wrote the generalized commutation relations for these
parafermionic algebras, and studied the representation theory
of the algebra of the $sl(3)_2/u(1)^2$ coset theory.
The theories corresponding to the $sl(3)_2/u(1)^2$, $sl(N)_2/u(1)^2$
cosets were known since a work by Gepner~\cite{Gepner:1987sm},
but the exact algebraic relations between the fields were never studied
in the past.

The results above demonstrate the power of algebraic approach
in the study of parafermionic conformal field theories.
The methods described will hopefully lead to
discovery of new algebras, and new two-dimensional conformal models.


\subsection*{Acknowledgment}

The author is grateful to Andrei Babichenko for his constant
interest and encouraging, discussions and comments on different
stages of the presented work.
The author also would like to thank Jim Lepowsky
for correspondence and explanations on Ref.~\cite{Dong_Lepowsky}.

An important part of the work was done during 2005, when the author
was a fellow within Marie Curie research training network
``Liegrits'' in Rome and Copenhagen, the author thanks the network
and its nodes for hospitality.

This research was supported by a Marie Curie Intra-European
Fellowships within the 6th European Community Framework Programme.





\begin{thebibliography}{99}




\bibitem{Fateev:1985mm}
V.~A.~Fateev and A.~B.~Zamolodchikov,
``Parafermionic Currents In The Two-Dimensional Conformal Quantum Field Theory
And Selfdual Critical Points In Z(N) Invariant Statistical Systems,''
Sov.\ Phys.\ JETP {\bf 62} (1985) 215
[Zh.\ Eksp.\ Teor.\ Fiz.\  {\bf 89} (1985) 380].

\bibitem{Zamolodchikov:1986gh}
  A.~B.~Zamolodchikov and V.~A.~Fateev,
  ``Disorder Fields In Two-Dimensional Conformal Quantum Field Theory And N=2
  Extended Supersymmetry,''
  Sov.\ Phys.\ JETP {\bf 63} (1986) 913
  [Zh.\ Eksp.\ Teor.\ Fiz.\  {\bf 90} (1986) 1553].

\bibitem{Fateev:1985ig}
  V.~A.~Fateev and A.~B.~Zamolodchikov,
  ``Representations Of The Algebra Of 'Parafermion Currents' Of Spin 4/3 In
  Two-Dimensional Conformal Field Theory. Minimal Models And The Tricritical
  Potts Z(3) Model,''
  Theor.\ Math.\ Phys.\  {\bf 71} (1987) 451
  [Teor.\ Mat.\ Fiz.\  {\bf 71} (1987) 163].

\bibitem{Gepner:1987sm}
  D.~Gepner,
  ``New Conformal Field Theories Associated With Lie Algebras And Their
  Partition Functions,''
  Nucl.\ Phys.\ B {\bf 290} (1987) 10.

\bibitem{Furlan:1992va}
  P.~Furlan, R.~R.~Paunov and I.~T.~Todorov,
  ``Extended U(1) conformal field theories and Z(k) parafermions,''
  Fortsch.\ Phys.\  {\bf 40} (1992) 211.

\bibitem{Camino:1998pd}
  J.~M.~Camino, A.~V.~Ramallo and J.~M.~Sanchez de Santos,
  ``Graded parafermions,''
  Nucl.\ Phys.\ B {\bf 530} (1998) 715,
  arXiv:hep-th/9805160.

\bibitem{Ding:2001ns}
  X.~M.~Ding, M.~D.~Gould and Y.~Z.~Zhang,
  ``Twisted parafermions,''
  Phys.\ Lett.\ B {\bf 530} (2002) 197
  [arXiv:hep-th/0110165].

\bibitem{Dotsenko:2002gs}
  V.~S.~Dotsenko, J.~L.~Jacobsen and R.~Santachiara,
  ``Parafermionic theory with the symmetry Z(5),''
  Nucl.\ Phys.\ B {\bf 656} (2003) 259,
  arXiv:hep-th/0212158.

\bibitem{Dotsenko:2003zc}
  V.~S.~Dotsenko, J.~L.~Jacobsen and S.~Raoul,
  ``Parafermionic theory with the symmetry Z(N), for N odd,''
  Nucl.\ Phys.\ B {\bf 664} (2003) 477,
  arXiv:hep-th/0303126.

\bibitem{Dotsenko:2003kg}
  V.~S.~Dotsenko, J.~L.~Jacobsen and R.~Santachiara,
  ``Conformal field theories with Z(N) and Lie algebra symmetries,''
  Phys.\ Lett.\ B {\bf 584} (2004) 186,
  arXiv:hep-th/0310102.

\bibitem{Dotsenko:2003ui}
  V.~S.~Dotsenko, J.~L.~Jacobsen and R.~Santachiara,
  ``Parafermionic theory with the symmetry Z(N), for N even,''
  Nucl.\ Phys.\ B {\bf 679} (2004) 464,
  arXiv:hep-th/0310131.

\bibitem{Dotsenko:2005se}
  V.~S.~Dotsenko and R.~Santachiara,
  ``The third parafermionic chiral algebra with the symmetry Z(3),''
  Phys.\ Lett.\ B {\bf 611} (2005) 189,
  arXiv:hep-th/0501128.

\bibitem{Jacob:2005kc}
  P.~Jacob and P.~Mathieu,
  ``A quasi-particle description of the M(3,p) models,''
  Nucl.\ Phys.\ B {\bf 733} (2006) 205,
  arXiv:hep-th/0506074.

\bibitem{Jacob:2005jz}
  P.~Jacob and P.~Mathieu,
  ``The Z(k)(su(2),3/2) parafermions,''
  Phys.\ Lett.\ B {\bf 627} (2005) 224,
  arXiv:hep-th/0506199.

\bibitem{Dong_Lepowsky}
C.~Dong and J.~Lepowsky,
``Generalized vertex algebras and relative vertex operators'',
Progress in Mathematics, 112, Birkh\"auser, Boston, 1993.

\bibitem{Bakalov_Kac}
B.~Bakalov and V.~Kac,
``Generalized Vertex Algebras'',
in ``Lie theory and its applications in physics VI,''
ed. V.K.~Dobrev et al., Heron Press, Sofia, 2006,
arXiv:math.QA/0602072 .

\bibitem{Belavin:1984vu}
  A.~A.~Belavin, A.~M.~Polyakov and A.~B.~Zamolodchikov,
  ``Infinite Conformal Symmetry In Two-Dimensional Quantum Field Theory,''
  Nucl.\ Phys.\ B {\bf 241} (1984) 333.

\bibitem{Ginsparg:1988ui}
P.~H.~Ginsparg,
``Applied Conformal Field Theory,''
 Champs, cordes et phenomenes critiques (Les Houches, 1988),  1--168,
North-Holland, Amsterdam, 1990,
arXiv:hep-th/9108028.

\bibitem{DiFrancesco:1997nk}
  P.~Di Francesco, P.~Mathieu and D.~Senechal,
  ``Conformal field theory,''
Graduate Texts in Contemporary Physics.
Springer-Verlag, New York, 1997.

\bibitem{Kac book}
V.~G.~Kac,
{``Vertex algebras for
beginners''}, Providence: AMS,
University Lecture Notes, Vol. 10, 1996, Second
edition, 1998.

\bibitem{Lepowsky_Li}
J.~Lepowsky and H.~Li,
``Introduction to Vertex Operator Algebras and Their Representations'',
Progress in Mathematics, 227, Birkh\"auser, Boston, 2004.


\bibitem{Ree}
R.~Ree,
``Generalized Lie elements''.
Canad.\ J.\ Math.\ {\bf 12} 1960 493–502.

\bibitem{Rittenberg:1978mr}
  V.~Rittenberg and D.~Wyler,
  ``Generalized Superalgebras,''
  Nucl.\ Phys.\ B {\bf 139} (1978) 189.

\bibitem{Rittenberg:1978df}
  V.~Rittenberg and D.~Wyler,
  ``Sequences Of Z(2) x Z(2) Graded Lie Algebras And Superalgebras,''
  J.\ Math.\ Phys.\  {\bf 19} (1978) 2193.

\bibitem{Scheunert:1978wn}
  M.~Scheunert,
  ``Generalized Lie Algebras,''
  J.\ Math.\ Phys.\  {\bf 20} (1979) 712.

\bibitem{Thielemans:1994er}
K.~Thielemans,
``An Algorithmic approach to operator product expansions, W algebras and W
strings,''
arXiv:hep-th/9506159.

\end{thebibliography}
\end{document}